# Coating-free Underwater Breathing via Biomimicry

Sankara Arunachalam[1,2*], Muhammad Subkhi Sadullah[2] & Himanshu Mishra[1,2,3*]

[1]Environmental Science and Engineering Program, Biological and Environmental Sciences and Engineering, Division, King Abdullah University of Science and Technology (KAUST), Thuwal 23955-6900, Kingdom of Saudi Arabia

[2]Water Desalination and Reuse Center, King Abdullah University of Science and Technology

[3]Center for Desert Agriculture, King Abdullah University of Science and Technology

[*]Corresponding authors: sankara.arunachalam@kaust.edu.sa & himanshu.mishra@kaust.edu.sa

**Abstract**

Numerous natural and engineering scenarios necessitate entrapment of air pockets or bubbles on submerged surfaces, e.g., aquatic insects, smartphones, and membranes for separation and purification. Current technologies for bubble entrapment rely heavily on perfluorocarbon coatings, which limits their sustainability and applications. Here, we investigate doubly reentrant cavities (DRCs) – a biomimetic microtexture capable of entrapping air under wetting liquids – under static and dynamic pressure cycling. The effects of positive (>1 atm), negative (< 1atm), and positive-negative cycles are studied across a range of pressure amplitudes, ramp rates, intercycle intervals, and water column heights. Remarkably, the fate of the trapped air under pressure cycling falls into the following three distinct regimes: the bubble (i) monotonically depletes, (ii) remains indefinitely stable, or (iii) starts growing. This hitherto unrealized richness of underwater bubble dynamics will guide the development of coating-free underwater technologies and provide clues into the curious lives of air-breather aquatic/marine insects.

When an air-breathing animal ventures under water, the robustness of the air entrapped on/around body is a matter of life and death. Broadly, the air entrapment takes place in the form of compressible bubbles or incompressible "plastron", i.e., physical gill[1]. Examples of air breathers exploring the underwater realm include, the aquatic beetle (*Potamodytes tuberosus*) that feeds on microbial slime under shallow rock surfaces in flowing waters[2]; the sea-skaters (e.g., *Halobates germanus*, *H. melleus*, and *H. hayanus*) that may accidentally drown after being hit by waves or rain drops and yet survive underwater for over 18 h[3-5]; and the semi-aquatic Anolis lizard species (e.g., *A. eugenegrahami* and *A. aquaticus*) that can respire for up to 18 minutes under water[6]. These animals have evolved strategies for combining their water-repellent secretions with specialized hairs or skin patterns to sustain underwater respiration[5-14]. Animal behavior under these life-threatening scenarios are fascinating but poorly understood; also, there is a growing expectation that their biomimicry will unleash sustainable technologies for frictional drag reduction[15-17], fouling-resistant membranes in fermenter broths[18], mitigating cavitation damage[19], packaging electronics[20,21], oxygen extraction in water[22,23], food-water security[24,25] and breathing textiles[26].

We are quite impressed by springtails (Collembola), ubiquitous arthropods that live in moist soils where flooding is frequent (Fig. 1A-C and Fig. S1). They breathe through their exoskeleton, which features cavities that have "mushroom-shaped" overhangs at their inlets, commonly known as doubly reentrant cavities (DRCs)[10,27-30]. The DRC architecture has been extensively investigated for enhancing boiling nucleation[31-34]. In their pioneering contribution, Kim & co-workers[32] inverted the DRC architecture to realize doubly reentrant pillar (DRP) microtexture that can repel droplets of wetting liquids of surface tensions as low as 10 mN/m (placed from the top)[16]. Curiously, the air-filled (or the Cassie-states)[35] on DRP arrays may catastrophically transition to the fully-filled (or the Wenzel state)[36] if the wetting liquid approaches laterally, e.g., at the array boundary or in the presence of localized defects/damage. The DRC architecture does not undergo such catastrophic wetting transitions due to its compartmentalized nature[28,37-44]. For example, recently, it was demonstrated for DRC submersion under hexadecane, where the intrinsic contact angle was $\theta_o = 20°$, that the entrapped air was intact even after a month[39]. In contrast, simple cylindrical cavities of the same size, and chemical make-up got filled in ~1s. This >10$^7$-times delay in wetting transitions underscores the efficacy of the DRC architecture toward realizing coating-free liquid repellence, which is desirable due to the adverse effects of perfluorinated chemicals[45,46]. Proof-of-concept demonstrations for DRCs–based gas-entrapping microtextured surfaces (GEMS)[47] and gas-entrapping membranes (GEMs)[48,49] from low-cost water-wet materials have appeared, which may unlock their potential for green technologies. Researchers have studied the dependence of the stability of the air-filled states on the liquid vapor pressure, condensation, apparent contact angles, and gas diffusion[39,50]. However, to fully assess the pros and cons of this approach, it is crucial to investigate the robustness of DRCs against pressure fluctuations – static and dynamic—that are common in practical scenarios, such as pipes and fittings[51], cavitation in pumps[52,53], microfluidic devices[54,55], pulse flows for cleaning membrane[56] and oscillating bubble induced jet flow for cleaning surfaces[57]. How the DRC air entrapment would fare under cyclic pressure under water, i.e., the longevity of Cassie-to-Wenzel transitions, is largely unknown and unexplored.

In this contribution, we submerge DRCs under varying depths of water columns and investigate the stability of the entrapped air on the pressurization/depressurization rates, the time intervals between the cycles, and the gas saturation level of the water. A broad spectrum of techniques are utilized to address the following interrelated elemental questions: (i) how does the breakthrough pressure vary with the rate of pressurization; (ii) how does the trapped air "bubbles" respond to cyclic pressure of positive ($>$ 1atm), negative ($<$ 1atm), and positive–negative triangular waves; (iii) what are the effects of the time intervals between the cycles on the bubble stability and DRC filling time; (iv) how do the wetting transitions depend on the water column height and the gas saturation level. Notably, our findings reveal that via DRCs it is possible to realize coating-free microtextured surfaces that can trap air indefinitely under cyclic pressure based on judicious selection of the cavity size, the water column height, and the time interval between the cycles.

**Results**

DRCs were realized on commercial $SiO_2$/Si wafers via microfabrication protocols that we have reported recently[37-39,47] (Fig. 1 D-G, see details in Methods). Briefly, photolithography, dry etching of $SiO_2$ and Si layers, and thermal oxide were exploited to microfabricate DRCs of diameter $D = 200$ µm but with two depths, $h \approx 55$ and $67$ µm. We realized single DRC per $\sim 1 \times 1 cm^2$ $SiO_2$/Si wafer substrate, instead of an array, because neighboring air bubbles in water[58,59] or evaporating droplets in the air[60] are known to exchange gas (or vapor) with each other and cause interference in determining the lifetime. Therefore, our samples with single DRCs obviated such an interference in the wetting transitions.

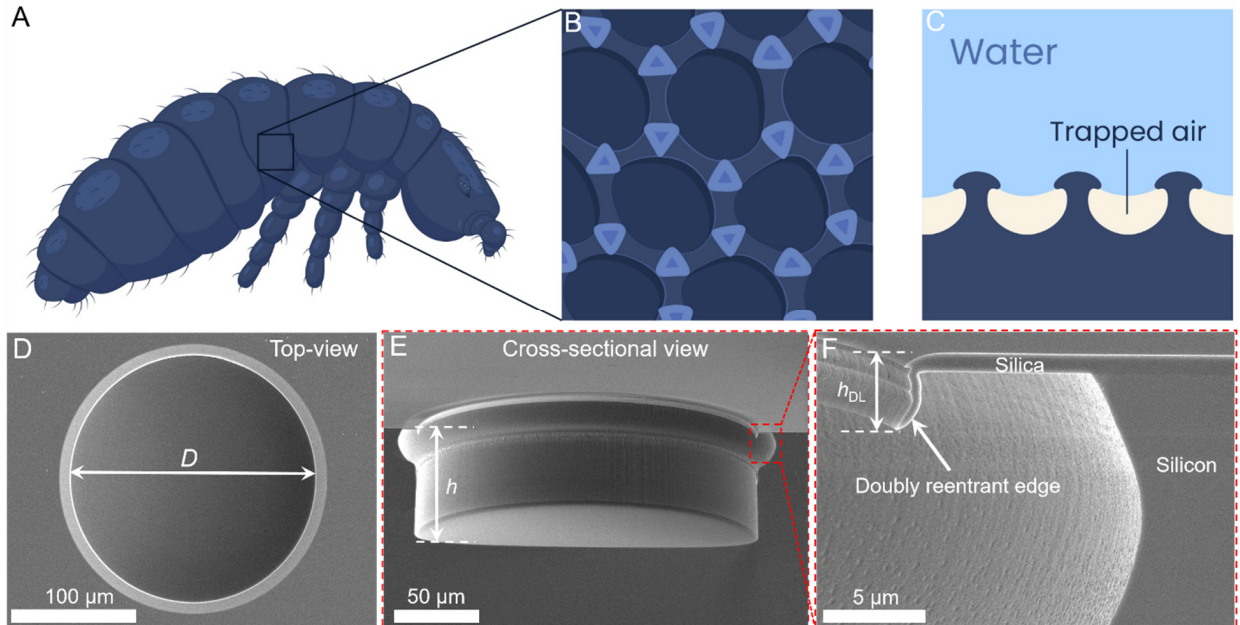

**Fig. 1: Illustration of springtail cuticles and representative scanning electron micrographs (SEMs) of doubly reentrant cavity**. (**A-C**) Schematics of cuticles of springtails (Collembola) with mushroom-shaped overhanging features enable them to entrap the air on immersion in water. Artistic reconstruction of the cuticle and its cross-sectional view inspired by Fig. S1 and images presented in ref.[28]. (**D-F**) Representative SEM of the top-view, the cross-sectional view, and the magnified view of DRCs of diameter, $D$ = 200 µm, and depth, $h \approx 67$ µm studied here.

Since a typical sample investigated in this work was ~1×1cm$^2$ in size with a DRC of diameter 200 µm in the middle, most of the sample surface was smooth and flat. Wetting of the flat silica surface was characterized, first, with water droplets of volume 6 µL by advancing and retracting them at ±0.2 µL/s. The advancing and receding apparent contact angles were, $\theta_A = 65° \pm 2°$ and $\theta_R = 30° \pm 2°$. However, standard contact angle goniometry could not be applied to characterize the wetting of individual DRC of size $D$ = 200 µm, and depth, $h \approx 67$ µm. In response, we applied the newest criterion for characterizing omniphobicity of surfaces via immersion[37]. Given silica's hydrophilicity, thermodynamics mandates that the DRC–water–air system must reside in the fully-filled (or the Wenzel) state[36,41]. However, as the sample was submerged under a 5 mm-high water column, the DRC entrapped air inside it and the subsequent Cassie-to-Wenzel wetting transition took >13 days (Fig. S2). This result underscores the efficacy of this coating-free approach under hydrostatic scenarios.

**Breakthrough pressure and effect of ramp rate**

Next, the mechanical stability of the air entrapped inside the submerged DRCs was challenged under a variety of hydrostatic pressurization and depressurization scenarios. We built a customized pressure cell that facilitated the samples' submergence under water columns of known heights while the headspace was pressurized at rates within the following range: 0.2 – 65 kPa/s (Figs. 2A and S3, Methods). The cell was fitted with white light microscope attached with camera to record wetting transitions from the top (Methods).

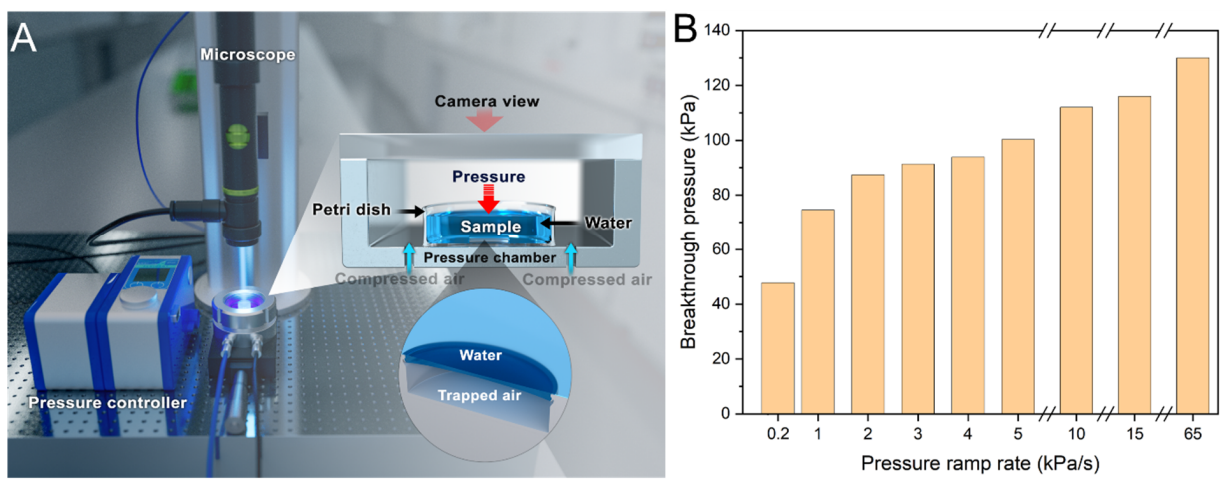

**Fig. 2: Experimental setup and influence of pressurization rate on breakthrough pressure.** **(A)** Schematic of the custom-built cell for studying static and cyclic pressure effects on DRCs. The headspace above the submerged sample is pressurized/depressurized and wetting transitions are recorded through an overhead optical microscope. **(B)** DRCs' breakthrough pressure (BtP) as a function of pressure ramp rate; the BtP is defined as the pressure at which the intruding water meniscus touches the cavity floor while being pinned at the DRC edge. The results showed that the higher the ramp rate, the higher the BtP. Note: the DRC's diameter was 200 µm and its depth was 55 µm.

We define the breakthrough pressure (BtP) for the DRC as the external hydrostatic pressure at which the water meniscus, pinned at the DRC edge, droops inside, and touches the cavity floor– the cavity "fails" (Fig. S4, Movie S1). Note: in this work, cavity "failure" is not the same as cavity "filling", where the latter ascribes to the fully-water-filled (Wenzel) state. BtP of a microtexture is a key descriptor of its robustness against liquid intrusion. Curiously, the BtP under 3 mm-thick water columns increased with the ramp rate within the abovementioned range. We observed BtP values to vary from 47–130 kPa, some 176% variation (Fig. 2B). For instance, the measured breakthrough pressures at the ramp rates of 0.2 kPa/s and 65 kPa/s were 47 kPa and 130 kPa, and it took 235 s and 2 s, respectively, to reach the BtP. Interestingly, the deeper the DRC for a given diameter, the higher was its BtP (Fig. S4). Note: the pressure values reported here and elsewhere correspond to the gauge pressure, i.e., the absolute pressure minus the atmospheric pressure, unless specifically noted. We explain these trends in the Discussion section.

**Continuous Cyclic Pressure (No Interval)**

In the next experiment, silica DRC samples were submerged under 3 mm-thick water columns and the headspace was subjected to cyclic pressure in the form of triangular waves separated by zero interval, $t_i = 0$ (Fig. 3A). The pressure amplitude ranged within 20–70 kPa, the rate of pressurization/depressurization was set at 1 kPa/s, and the fate of the air trapped inside the DRCs was recorded. Figure 3B presents the representative behavior of the air-water interface under 40 kPa cyclic pressure (Movie S2). Since the pressure amplitude was lower than the breakthrough pressure, water remained pinned at the DR edge while the water–vapor interface drooped inside the cavity, but it did not touch the cavity floor. In this process, the interfacial curvature of the air–water interface changed from being flat to concave with the increasing pressure. Pinning at the DRC edge was confirmed via confocal laser scanning microscopy (Figs. S5, S6 and Section S1 in the SI). On full depressurization, however, the interface did not recover to its original position, and following each cycle it kept on drooping into the DRC and eventually touched the cavity floor (after 6 cycles). Here, we define the "failure cycle" as the number of the pressure cycle at which the cavity floor was touched (Fig. 3 B-V). With the increasing pressure amplitude (ramp rate: 1kPa/s) the DRCs failed sooner, i.e., at a lower cycle number (Fig. 3C). We explain these trends in the Discussion section.

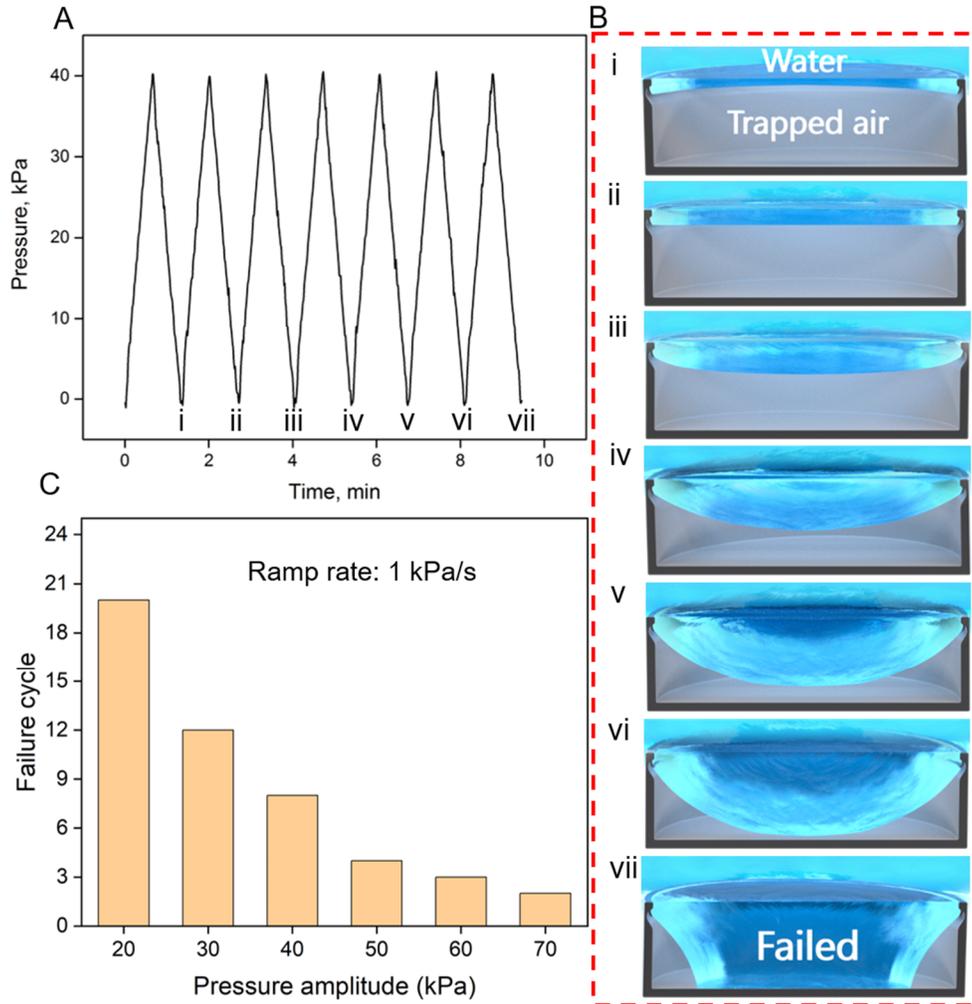

**Fig. 3: Submerged DRCs under continuous cyclic pressure ($t_i = 0$).** (**A**) 40 kPa triangular pressure cycles with zero interval time. (**B**) Schematic illustrates qualitatively how the interface progresses inside the cavity during each cycle and eventually "fails" i.e., touches the cavity floor, at the 7$^{th}$ cycle. (**C**) Keeping the fixed ramp rate of 1kPa/s, as the pressure amplitude increases the failure cycle# decreases.

**Cyclic Pressure with Intervals**

Next, we modified the above-described experiment by simply separating the triangular pressure waves of 40 kPa amplitude by time intervals, $t_i$, of 0, 1, 2, 3, 4, 5, 6, 8, or 10 min (Fig. 4 inset). The effect was profound – depending on the $t_i$, the cavity dimensions, the water column height, two extreme regimes emerged: (i) **the unstable regime**: wherein, $t_i$ did not significantly influence the failure cycle# for the DRC and they failed within a few cycles (see data for $t_i \leq 5$ min in Fig. 4, Movie S3); and (ii) **the stable regime**: wherein $t_i$ delayed the cavity failure indefinitely (see data for $t_i \geq 6$ min in Fig. 4, Movie S3). Specifically, for $t_i \geq 6$ min, the air trapped inside the DRCs became immune to depletion – pinned at the DRC edge, the air–water interface flopped up and down with the cyclic pressure for over 400 cycles (48 hr) and did not deplete at all; after this, the experiments were discontinued. Also, after pressure cycling in the stable regime, the BtP was found to be significantly higher, e.g., after 700 cycles of 40 kPa triangular waves with $t_i$ =10 min, the BtP had increased by 28% (Fig. S7). To our knowledge, such a behavior of DRCs has never been reported before, and we explain the underlying mechanisms in the discussion section below.

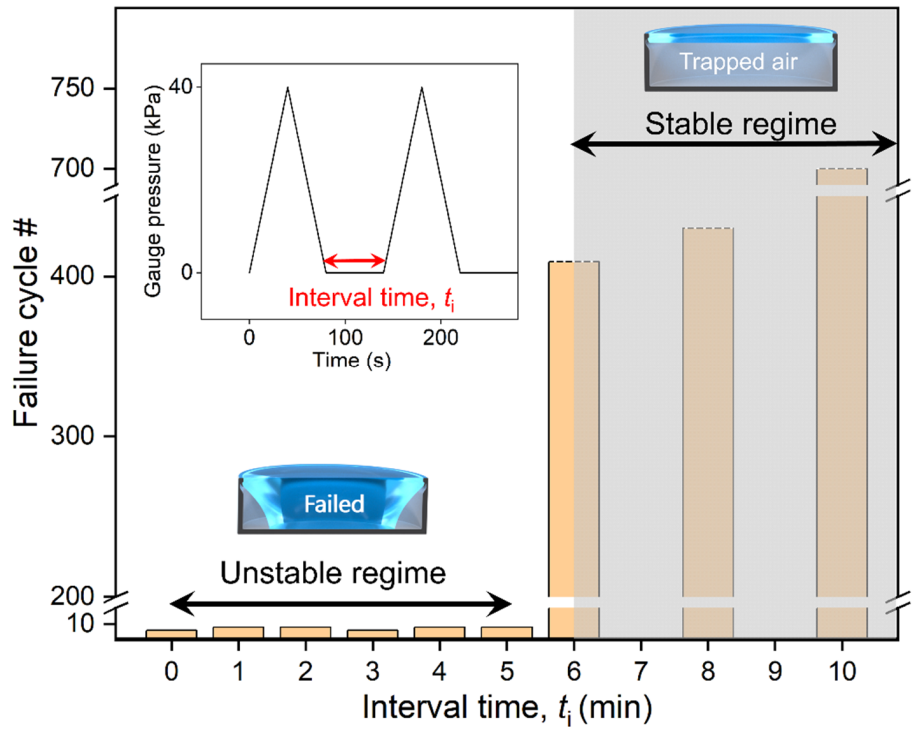

**Fig. 4: Submerged DRCs under cyclic pressure with intervals ($t_i$ > 0).** Dramatic effects of the time interval, $t_i$, during 40 kPa triangular pressure cycling on the fate of the air trapped inside DRC submerged under 3 mm-thick water column. For $t_i$ ≤5 min, the DRCs failed within 8 cycles. In contrast, for $t_i$ ≥ 6 min, the DRCs became immune to air depletion – the air was intact even after > 400 cycles, after which the experiments were discontinued.

**Positive, Negative, and Alternating Positive–Negative Pressure Cycles**

We broadened our investigation to three pressure cycling scenarios relevant in natural and applied contexts (Fig. 5A–C). They were: (i) positive pressure cycling (Fig. 5A), similar to human breathing with continuous positive airway pressure (CPAP)[61]; (ii) negative pressure cycling (Fig. 5B), i.e., reducing the pressure below the ambient pressure and (iii) alternating positive–negative cycles(Fig. 5C) similar to bubbles in acoustic fields, and fluctuating flows over surfaces[51,53,56,62].

For the CPAP-type cycling, the breathe-in pressure was set to 8 kPa, the breathe-out pressure was set to 3.5 kPa, the breathing duration (in/out) was set to 1.5 sec, and the time interval between each cycle was 0 s (Fig. 5A, Movie S4). The following two water column heights were tested, $H = 2$ and 6 mm. For the 6 mm water column, the DRC failed at 36 min (720 cycles). Remarkably, when the water column height was reduced to $H = 2$ mm, keeping the other experimental parameters the same, the cavity failure was prevented for over 40 hrs (48,000 cycles), after which the experiment was discontinued. We explain the physics behind this 6600% enhancement in the Discussion section below. Also, we conducted a BtP test after 40 hrs of cycling and found it to be 87 kPa, which was 16% higher than that right after immersing the DRC in water.

For the second scenario, DRCs were immersed in 3 mm-thick water columns and subjected to negative triangular pressure cycling of -40 kPa amplitude via a vacuum suction device (the ramp rate was ±1kPa/s and $t_i = 0$) (Fig. 5B). During each cycle, the pressure of the headspace (Fig. 2A) was reduced from $P = 101$ kPa to $P = 61$ kPa. To the trapped air inside the DRC, this pressure change was equivalent to falling from $P = 0$ to $P = -40$ kPa (Fig. S8). Under these conditions, the trapped bubble expanded and shrank, and with the increasing number of cycles, it grew out of the cavity (Fig. 5B, D, and Movie S5). At $t = 33.3$ hrs and $P = 0$ kPa, the diameter of the bubble became 450 µm, even though it was still pinned at the mouth of the DRC. We induced mechanical vibrations to check the mechanical stability of the interface, i.e., whether the DRC would get catastrophically filled if a significant portion of the bubble departed. Remarkably, as a major portion of the bubble departed, the DRCs recovered its initial Cassie state (Fig. 5D, Movie S6). At this point, the BtP, measured at the ramp rate of 1 kPa/s, was ~75 kPa.

Lastly, the headspace above the DRCs immersed in 3 mm-thick water columns was subjected to alternating positive–negative pressure cycles of amplitude ±40 kPa (at the ramp rate of ±1kPa/s and $t_i = 0$) via a combination of positive air pressure and vacuum suction to realize (Figs. 2A and 5C). Intuitively, one would expect that if the same pressure amplitude is applied above and below the baseline, the overall effect on the bubble size should be null. Counterintuitively, the bubble grew in its size over time (Movie S7). We induced mechanical vibrations to detach the enlarged bubble from the cavity; and after its departure, the DRCs recovered its initial Cassie state.

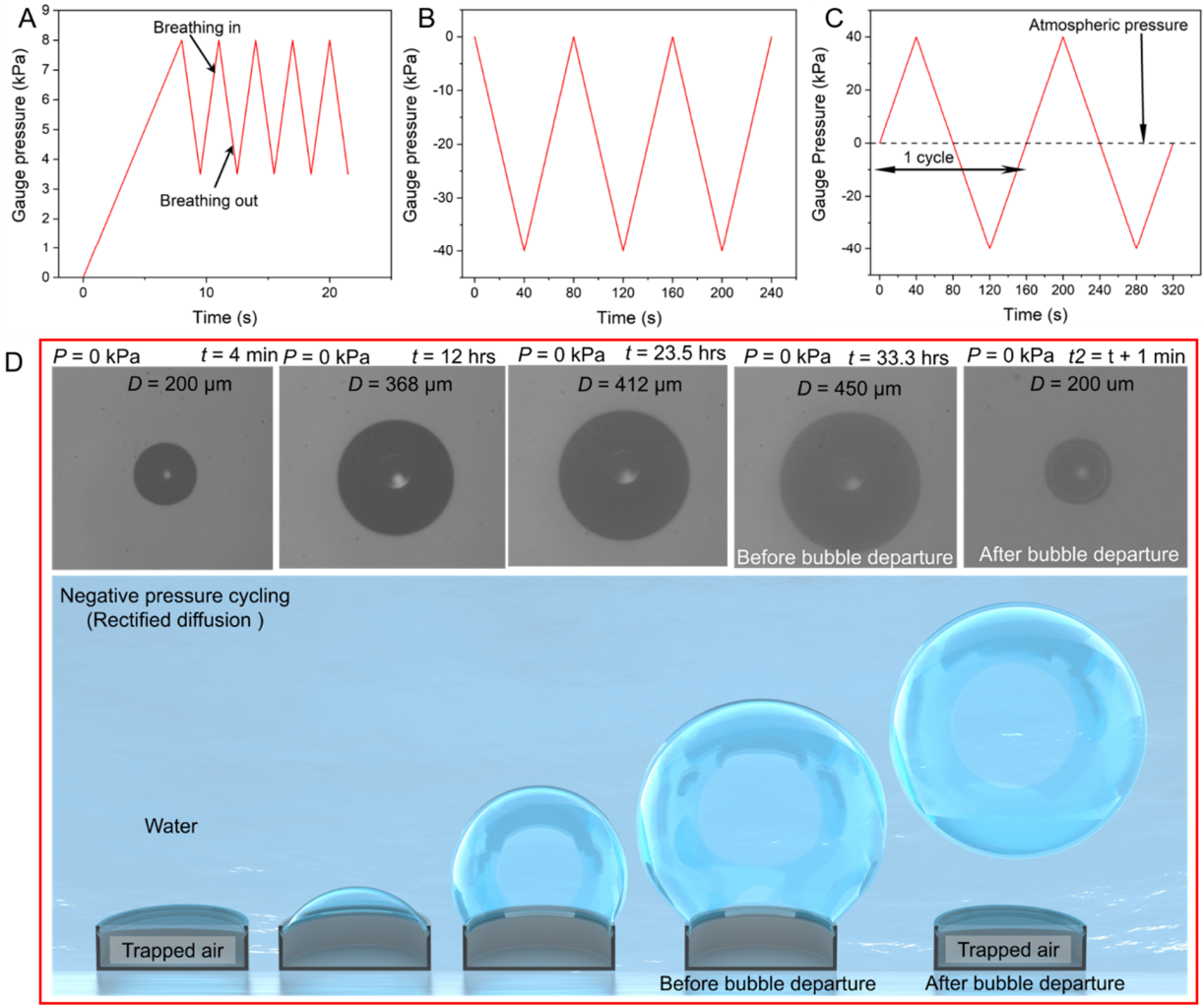

**Fig. 5**: **Submerged DRCs subjected to positive (> 1 atm), negative (< 1 atm), and alternating positive–negative (≳1 atm) pressure cycles**. (**A**) Positive pressure cycles akin to basal human breathing. (**B**) Negative pressure cycles. (**C**) Positive and negative and pressure cycling. (**D**) The bubble diameter under continuous negative cycles for 33 hrs increased to 450 µm, even though it remained pinned at the DRC mouth, and it exhibited a stable interface even after bubble rupture due to mechanical vibration. The top panel shows the bubble growth's optical image sequence (top-view), and the bottom panel shows the schematics. Water column height: 3 mm.

## Discussion

Here, we draw together the results and discuss the underlying factors and mechanisms. First, we discuss the effects of the pressure ramp rates on the BtP (Fig. 2B). As the headspace is pressurized, water tends to enter the cavity and compresses the entrapped air. Over time, this air dissolves into the water column and transports via diffusion, creating space for the water to enter/occupy (Fig 6A-C). Therefore, a slow ramp rate facilitates easier liquid penetration and yields a low BtP. In contrast, if the ramp rate is high, the dissolution of the trapped air into the water is minimal, and the liquid penetration is resisted by the gas compression which raises the BtP. We used the ideal gas equation[63] to describe the predict the effect of applied gauge pressure $P$ on the gas volume inside the cavity $V$ assuming isothermal conditions:

$$V = V_i \frac{P_i}{P_i+P}, \tag{1}$$

where $V_i$ and $P_i$ are the initial gas volume and the initial external pressure. $P_i$ is defined as $P_i = (P_{atm} + \rho g H - P_v)$ which accounts for the atmospheric $P_{atm}$, hydrostatic $\rho g H$, and vapor pressures $P_v$.

To account the effect of diffusion, we employed the Fick's first law along with the Henry's law. According to the Henry's law, with the increase in the gas pressure, the solubility of the gas in water increases. Considering the air to be a 79:21 mixture of $N_2$ and $O_2$ gases, whose Henry's constants are known, the equilibrium concentration of the dissolved air in the water ($c_\infty$) in equilibrium with the atmosphere at 1 atm pressure is expected to be, $c_\infty = 23$ g/m$^3$ [64,65]. Now, as the water is pressurized, the new equilibrium gas concentration at the air-water interface increases, which can be calculated using the Henry's law as, $c_0 > c_\infty$. This leads to a buildup of a vertical concentration gradient between the interface and the bulk, $c_0 - c_\infty$, which drives gas dissolution as per the Fick's first law, $J = -D_G \frac{\Delta c}{l}$, where $J$ is the gas flux, $D_G$ is the diffusion constant, $l$ is diffusion length. Assuming that the trapped air exhibits the ideal gas behavior ($p_G V = nRT$), and the mass flux of the air from the cavity to the water per unit area $[J = \left(\frac{dn}{dt}\right)/A(t)]$, where $p_G = P_i + P$, $V$ are pressure and volume of the trapped air, $n$ is the number of air moles in the trapped air, $R$ is the universal gas constant, $T$ is temperature, $A$ is the gas-liquid interface area. Based on the above equations, we can rewrite the Fick's first law as (Section S2 in the SI)[66]

$$\frac{d(p_G(t)V(t))}{dt} = -RTA(t)\frac{D_G}{K_H}\frac{(p_G(t)-s \cdot p_{G,0})}{l}, \tag{2}$$

where $K_H$, $l$, and $s$ are respectively the Henry's constant, the diffusion length, and the gas solubility. The pressure inside the cavity is denoted by $p_G$, while $p_{G,0}$ is the gas partial pressure at atmospheric pressure.

Eq. (1) describes the effect of pressure change to the trapped gas volume, while Eq. (2) describes the gas volume loss due to diffusion. By numerically solving Eq. (1) and (2) iteratively for an increasing $P$, we obtained the BtP for various pressure ramp rates as well as the failing cycles for various pressure amplitudes. Figures 6D and 6E show the comparison of the experimental results with the theoretical results, which demonstrates that the model can successfully capture the physics underlying the experimental observations. The detail of how the simulation was carried out is explained in the Supplementary Information (Section S2).

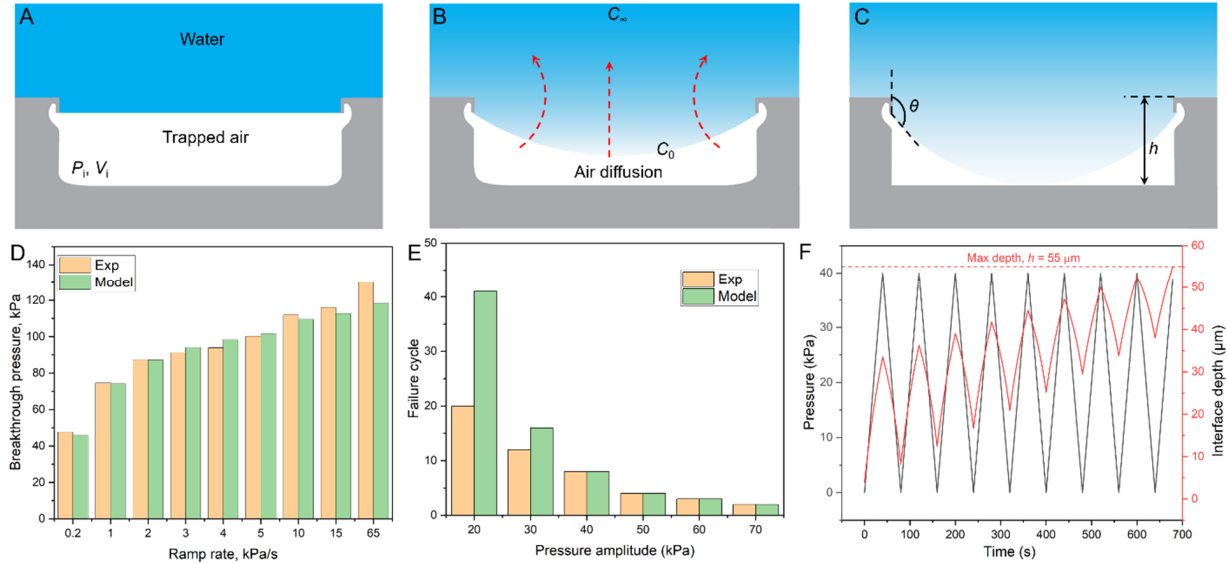

**Fig. 6: Predicted breakthrough pressures and failure cycles.** (A-C) Schematics of the air–water interface at/inside DRC with respect to increasing pressure. (D) Comparison of the predicted breakthrough pressures with the experimentally observed values for submerged DRCs as a function of the pressure ramp rate. (E) Comparison of the predicted failure cycles with the experimentally observed values for submerged DRCs as a function of pressure amplitude. (F) Predicted distance between the DR edge and the liquid-vapor interface depth (which we refer to as interface depth) for 40 kPa cyclic pressure.

Increasing the ramp rate gave minimal time for gas diffusion, hence higher BtP. Here, we also point out that BtP of common rough surfaces/membranes typically depends on the Laplace pressure, $P_L$, arising due to the curvature of the air–water interface at the microtexture's inlet. We estimate the maximum possible contribution of $P_L$ for our DRC using the expression, $P_L = -4\gamma \cos\theta_{max}/D$. Interestingly, $P_L \sim 1$ kPa was insignificant in comparison with the air pressure inside the DRC generated due to its compression, which underscores the importance of the entrapped gas and its time-dependent loss on BtP.

The effects of the pressure ramp rates on the BtPs is relevant to the underwater breathing insects, e.g. the alkali fly (*Ephydra hians*) that can crawl to 4–8 m deep waters for foraging and laying eggs[12]. It could maximize plastron respiration by diving deep rapidly to minimize the loss of the trapped gas, or simply stay in the shallow water where the hydrostatic pressure is low and wetting transitions would be slow. A case-by-case assessment of the various animals is warranted for biomimicry.

Next, we explain the emergence of the unstable and indefinitely stable regimes under pressure cycling and how they depend on $t_i$, the water column height, and the cavity dimensions (Figs. 3 and 4). Here, we need to decouple first the contributions of the dissolution and diffusion of the trapped air and the headspace air into the water column during pressurization and depressurization. During continuous pressure cycling or cycles separated by $t_i \leq 5$ min for the DRCs studied herein, the trapped air is lost via diffusion into water due to its pressure-induced undersaturation. Hence, the liquid meniscus approaches the cavity floor monotonically and the air entrapment falls in the unstable regime. The gradient of dissolved gas concentration in undersaturated water $(c_o - c_\infty) > 0$ results in continuous outward mass flux ($J < 0$), until trapped air dissolves in water. Using Eq. (1) and (2) but for cyclic variation of $P$, we capture the failure cycle due to cyclic pressure with different pressure amplitudes (Fig 6E). Fig 6F shows the failure cycle for 40 kPa amplitude cycle and the corresponding distance between the DR edge and the liquid-vapor interface depth, which we refer to as interface depth. It reveals that the interface depth increases for each cycle, indicating that the gas volume is lost through diffusion at each cycle. In contrast, the indefinitely stable regime for the trapped air for the DRCs studied in this work is observed when the cycles are separated by $t_i \geq 6$ min (Fig. 4). This interval time enables the water column to get saturated under the applied pressure via air diffusion from the headspace. Which results in $(c_o - c_\infty) = 0$, hence no mass flux ($J = 0$), and therefore no cavity failure. In our experiments, as well as for insects under shallow waters[2], the gas exchange at the headspace controls the saturation level of the water, which dictates whether the bubble will "breathe" outward or inward. The time dependence of the spatial gas concentration is governed by the Fick's second law, $\frac{\partial c}{\partial t} = D_G \frac{\partial^2 c}{\partial z^2}$.[64,66,67] Considering I-D diffusion for simplicity, the time required to saturate the entire water column height can be estimated by the equation, $t_D \approx \frac{z^2}{2D}$,[64] for 3 mm water column, it may take $t_D \approx$ 38 minutes to saturate the water under hydrostatic conditions. This means that for the air trapped inside the DRC under a liquid column height, $H$, and subjected to pressure cycling, if the interval time, $t_i$, is such that the cavity failure cycle time is greater than $t_D$, then the water column will get saturated before the water touches the DRC floor. In such a scenario, the curvature of the saturated air–water interface will diffuse air into the cavity until its curvature is lost i.e., it flattens. Since our experiments entail pressure cycling, the actual time to saturate the water column was longer (positive pressure cycles, Fig. 4), i.e., < 60 min. To test this, we varied the water column height from 3 mm to 6 mm and subjected to a 40 kPa cyclic pressure (ramp rate 1 kPa/s and $t_i = 6$ min). While for 3 mm column height, the system acquired the stable state, for 6 mm the DRC failed at the 7$^{th}$ cycle. If we reduced the water column height to 2 mm, then the stable regime could be reached under at $t_i = 5$ min under 40 kPa cyclic pressure. This underscores the importance of the DRC (or bubble) size, $t_i$, and $t_D$ to achieve water column saturation and realize breathing interfaces. This also provides clues for the bubble stability in the case of the aquatic beetle (*P. tuberosus*) that lives in shallow flowing waters, which are known to be saturated with air[2]. Figure 7 A-B summarizes the behaviors of trapped bubble under positive cyclic pressures with zero and non-zero $t_i$. According to Henry's law, pressurizing the water above the atmospheric pressure makes it undersaturated $(c_o - c_\infty) > 0$, with respect to the gas bubble; and depressurizing it below the atmosphere pressure results in supersaturation $(c_o - c_\infty) < 0$ with respect to the bubble (Fig. 7C, discussed below). For positive cyclic pressures, a careful selection of $t_i$ and $t_D$ may unlock the stable regime, where the bubble may exhibit indefinite stability. In this study, we utilized a vertical overhang length ($h_{DL}$) of 4 µm, which accounts for approximately 7% of the cavity depth ($h$ =55 µm). Our model predicts that increasing $h_{DL}$, while keeping the cavity depth constant, leads to a decrease in the breakthrough pressure and the failure cycle number (Fig. S9). Also, in this study, we utilized shallower cavities ( $h/D$ <1) to ensure that the liquid-vapor interface remain pinned at the DR edge. As the pressure increases, the curvature of the spheroid-shaped water-vapor interface keeps increasing until the water touches the cavity floor. However, depinning mode of failure[44,68] can be expected for the deeper cavities ( $h/D$ >1) where the increase in the hydrostatic pressure causes the primary meniscus to reach its maximum curvature ($\approx D/2$) and depinnes from DR edge onto the walls of the cavity. Thereafter, the advantages of DR features were lost, and DRCs behaved as simple cavities[50].

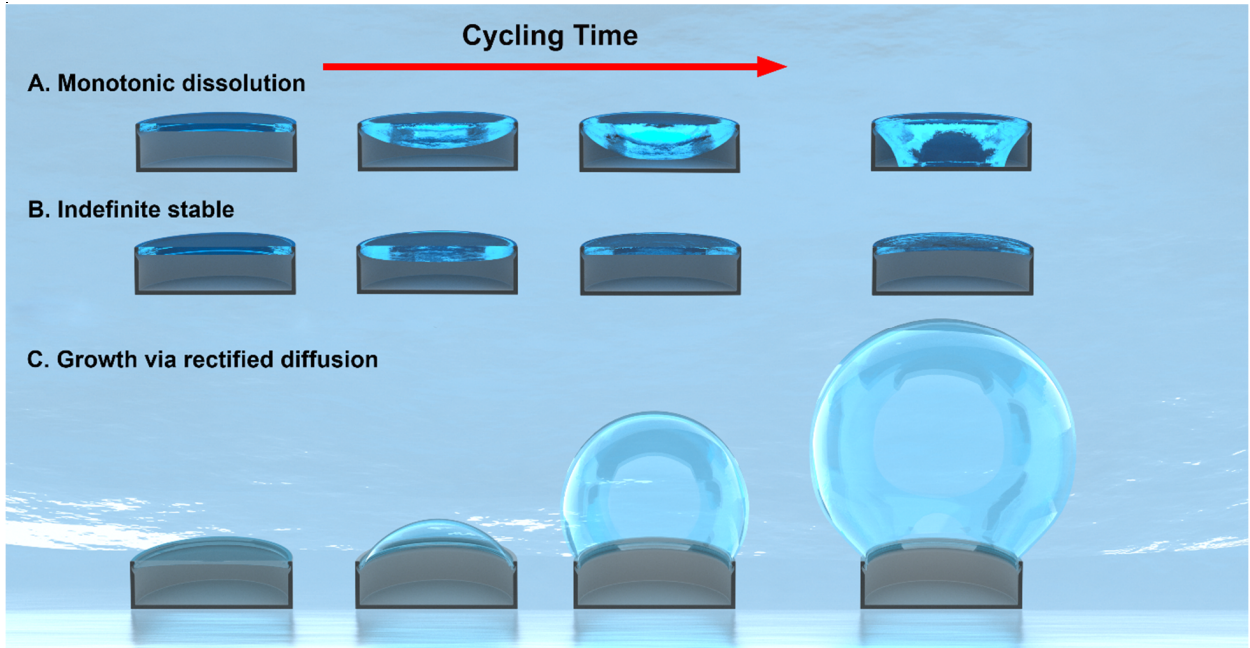

**Fig. 7: Summary of trapped bubble stability under cyclic pressure.** (**A**) Trapped air monotonically depletes due to diffusion under continuous positive cycles in undersaturated water. (**B**) Trapped air is indefinitely stable under positive cycling in saturated water. (**C**) Trapped bubble starts growing due to rectified diffusion under continuous negative, positive-negative cycles.

Next, we explain the curious observation of bubble growth under continuous negative and positive-negative pressure cycling, i.e., $t_i = 0$ (Fig. 5B–C). While it is intuitive for the purely negative cycles, it is not so obvious for the positive–negative pressure cycling as one would expect the bubble size to remain unchanged. In a typical cycle, as the pressure in the headspace drops, the water column becomes oversaturated, which transfers air into the entrapped gas pocket (as well as into the headspace). In oversaturated water, $(c_o - c_\infty) < 0$ results in inward mass flux ($J > 0$). Curiously, previous research has revealed that when a bubble is exposed to pressure cycling within a supersaturated liquid, it may start growing if the pressure amplitude is beyond a certain threshold – a phenomenon known as rectified diffusion[69,70]. While a detailed investigation of this process is beyond the scope of this study., scientific literature suggests that it's due to the non-linear combination of the enhanced surface area during expansion (area effect) and the "shell effect" that refers to the difference in the thicknesses of the liquid–gas mass transfer boundary during expansion and compression[71]. During compression, this boundary layer is thicker than that during expansion, which promotes more gas to diffuse in the bubble during its expansion than its loss during compression. For an insect respiring via plastron under water, it is therefore advantageous to be in shallow flowing water, because (i) it is saturated with oxygen, and (ii) during exhalation, as the plastron expands, it can register a net gain of $O_2$ gas[2]. More work is needed to ascertain whether the pressure amplitudes (positive/negative/positive-negative) and frequency of breathing are adequate for rectified diffusion[71].

Lastly, we explain the significantly higher breakthrough pressures recorded after pressure cycling experiments. This is related to the gas oversaturation of the water during positive pressure cycling, such that the system enters into the stable regime (Fig. 4). Now, if a BtP measurement follows, as water is pushed into the cavity, the gas from the cavity dissolves much more slowly into the water due to its oversaturation, which manifests as higher breakthrough pressure (Fig. S7).

**Conclusion**

To our knowledge, this is the first report on the effects of pressure cycling on submerged DRCs, and the findings are obviously relevant to hydrophobic and superhydrophobic surfaces/membranes that serve many practical applications. Prior to this report, DRCs immersed in water were known to monotonically lose the entrapped air due to the coupling between capillary condensation and gas diffusion[39,72]. Current results unveil a range of hitherto unrealized possibilities for the air trapped inside submerged DRCs under pressure cycling, viz. enhanced BtP, indefinitely long Cassie-states, rectified diffusion, and the recovery of the original Cassie-stater after losing the bubble, depending on the pressurization/depressurization rates, $t_i$, $t_D$, the water column height, and the DRC volume. These results may also provide clues to the understanding of air breathers' underwater vis-a-vis their surface microtexture, depths and durations explored, diving speeds, plastron stability under continuous and discontinuous breathing cycles[73,74], and rectified diffusion. Our toy model system may also facilitate the simulation of pressure cycling for human divers safety, which is otherwise studied via acoustic bubble growth at low frequencies[75]. Under the constraints of present and future regulations on perfluoroalkylsubstances[45,46], DRCs may afford a coating-free alternative for applications that require omniphobicity. Proof-of-concept demonstrations for DRCs have appeared recently under hydrostatic scenarios, such as for mitigating cavitation erosion[19], enhance heat transfer[34,76], microfluidics[77], separation and purification technologies[48,78], frictional drag reduction[15,79] oscillating-bubble powered microswimmer[80] . We recommend rigorous pressure cycling assessments under lower surface tension liquids as well as studying arrays of DRCs.

**Methods**

*Microfabrication:* We used layout designer (Tanner L- edit, V15.0) to create the cavity pattern and using a direct writer (Heidelberg Instruments, µPG501) exposed the design to 1.6 µm thick AZ5214E spin-coated SiO$_2$/ Si wafer (Silicon Valley Microelectronics, orientation 100, Single side polished, 4-inch diameter, 500 µm thickness, 2 µm thick oxide layer). The UV-exposed photoresist was removed in a bath of AZ-726 developer. The SiO2 and Si layers were etched using inductively coupled plasma (ICP) reactive-ion etching (RIE) and Deep ICP−RIE (Oxford instruments), respectively. In between these etching steps, 500 nm thick thermal oxide was grown over the etched wafer, using a Tystar furnace system to realize a doubly reentrant edge. After etching to the required depth the samples were cleaned in piranha solution ($H_2SO_4/H_2O_2$ = 3:1 by volume) at $T$ = 115 °C for 10 min, and spin-dried under a nitrogen environment. The wafer was then stored in a glass petri dish and placed inside a clean vacuum oven maintained at $T$ = 50 °C and gauge pressure $P_{vac}$ = -80 kPa for 5 days, after which the intrinsic contact angle of the silica layer stabilized to $\theta_o \approx 63°$. We adopted this microfabrication procedure from our previous work [37,39,47]

*Scanning electron microscopy (SEM):* We used Quattro (from Thermo Fisher Scientific) for SEM imaging. Microtextured samples were mounted on the stub using double-sided carbon tape, and prior to SEM, samples were coated with 3 nm of Iridium using a sputter coater (Quorum Q150T S) to eliminate charging effects.

*Contact angle measurement:* Apparent contact angles of water on the samples were measured using the Kruss Drop Shape Analyzer (DSA100E). The liquid was advanced/receded at the rate of 0.2 µL.s$^{-1}$. All the data were analyzed using the *Advance* software. The reported data points are an average of at least three measurements.

*Breakthrough pressure and cyclic pressure experiment*: We fixed the sample into the petri dish (35 mm petri dish from Celltreat) using permanent double-sided tape (scotch) and kept it inside the custom-built pressure cell. Before the experiment, we added the water ($\approx$ 3 ml at $\approx$ 0.3 ml/s rate) using a syringe and needle to submerge the sample and closed the pressure cell with a top cover. We used a fluigent pressure controller (Push and Push & Pull type), Keller (LEO3 digital manometer) pressure gauge, and compressed air/vacuum to increase/decrease headspace pressure. The required pressure ramp rate and cyclic pressure profile was programmed using microfluidic automation software provided by a Fluigent. The default pressure ramp rate was set as 1 kPa/s unless specified. Wetting transitions inside the cavity were observed from the top using monochrome digital camera (Edmund Optics, EO-5310) attached to Qioptiq objective, with a working distance of 9.5 cm. We used a screen recorder (VSDC) to continuously capture the ensuing wetting transitions and later analyzed them to extract breakthrough pressure, fail time, and failing cycle details. Unless stated otherwise, all the pressure value reported here corresponds to gauge pressure.

*Numerical calculations*: The numerical simulations in this study were performed using Mathematica software. The details about the algorithm and parameter values are provided in Section S4 in the Supplementary Information.

**References**

1. Seymour, R. S. & Matthews, P. G. Physical gills in diving insects and spiders: theory and experiment. *Journal of Experimental Biology* **216**, 164-170, (2013).
2. Stride, G. O. The respiratory bubble of the aquatic beetle, Potamodytes tuberosus, Hinton. *Nature* **171**, 885-886, (1953).
3. Cheng, L. Marine and freshwater skaters: differences in surface fine structures. *Nature* **242**, 132, (1973).
4. Mahadik, G. *et al.* Superhydrophobicity and size reduction enabled Halobates (Insecta: Heteroptera, Gerridae) to colonize the open ocean. *Scientific Reports* **10**, 1-12, (2020).
5. Cheng, L. & Mishra, H. Why did only one genus of insects, Halobates, take to the high seas. *PLoS Biol* **20**, e3001570, (2022).
6. Boccia, C. K. *et al.* Repeated evolution of underwater rebreathing in diving Anolis lizards. *Current Biology*, (2021).
7. Bush, J. W., Hu, D. L. & Prakash, M. The integument of water-walking arthropods: form and function. *Advances in insect physiology* **34**, 117-192, (2007).
8. Seymour, R. S., Jones, K. K. & Hetz, S. K. Respiratory function of the plastron in the aquatic bug Aphelocheirus aestivalis (Hemiptera, Aphelocheiridae). *Journal of Experimental Biology* **218**, 2840-2846, (2015).
9. Flynn, M. R. & Bush, J. W. Underwater breathing: the mechanics of plastron respiration. *Journal of Fluid Mechanics* **608**, 275-296, (2008).
10. Hopkin, S. P. *Biology of the springtails:(Insecta: Collembola)*.  (OUP Oxford, 1997).
11. Andersen, N. M. & Cheng, L. The marine insect Halobates (Heteroptera: Gerridae): biology, adaptations, distribution, and phylogeny. *Oceanography and marine biology: an annual review* **42**, 119-180, (2004).
12. van Breugel, F. & Dickinson, M. H. Superhydrophobic diving flies (Ephydra hians) and the hypersaline waters of Mono Lake. *Proceedings of the National Academy of Sciences* **114**, 13483-13488, (2017).
13. Gorb, S. N. *Functional Surfaces in Biology: Little Structures with Big Effects Volume 1*. Vol. 1 (Springer Science & Business Media, 2009).
14. Matthews, P. G. & Seymour, R. S. Diving insects boost their buoyancy bubbles. *Nature* **441**, 171-171, (2006).
15. Vakarelski, I. U. *et al.* Self-determined shapes and velocities of giant near-zero drag gas cavities. *Science Advances* **3**, (2017).
16. Liu, T. Y. & Kim, C. J. Turning a surface superrepellent even to completely wetting liquids. *Science* **346**, 1096-1100, (2014).
17. Park, H., Choi, C.-H. & Kim, C.-J. Superhydrophobic drag reduction in turbulent flows: a critical review. *Experiments in Fluids* **62**, 1-29, (2021).
18. Stephanopoulos, G. Challenges in engineering microbes for biofuels production. *Science* **315**, 801-804, (2007).
19. Gonzalez-Avila, S. R. *et al.* Mitigating cavitation erosion using biomimetic gas-entrapping microtextured surfaces (GEMS). *Science Advances* **6**, eaax6192, (2020).
20. Yu, Q., Xiong, R., Li, C. & Pecht, M. G. Water-resistant smartphone technologies. *IEEE Access* **7**, 42757-42773, (2019).
21. Baylakoğlu, İ. *et al.* The detrimental effects of water on electronic devices. *e-Prime-Advances in Electrical Engineering, Electronics and Energy* **1**, 100016, (2021).

22  Lee, J., Heo, P. W. & Kim, T. Theoretical model and experimental validation for underwater oxygen extraction for realizing artificial gills. *Sensors and Actuators A: Physical* **284**, 103-111, (2018).
23  Atherton, S. *et al.* Plastron Respiration Using Commercial Fabrics. *Materials* **7**, 484-495, (2014).
24  Odokonyero, K., Gallo Jr, A. & Mishra, H. Nature-inspired wax-coated jute bags for reducing post-harvest storage losses. *Scientific Reports* **11**, 15354, (2021).
25  Gallo Jr, A. *et al.* Nature-inspired superhydrophobic sand mulches increase agricultural productivity and water-use efficiency in arid regions. *ACS Agricultural Science & Technology* **2**, 276-288, (2022).
26  Ebnesajjad, S. *Expanded PTFE applications handbook: Technology, manufacturing and applications*.  (William Andrew, 2016).
27  Hensel, R. *Robust omniphobic surfaces by mimicking the springtail skin morphology*, Technische Universität Dresden, (2014).
28  Helbig, R., Nickerl, J., Neinhuis, C. & Werner, C. Smart Skin Patterns Protect Springtails. *PLOS ONE* **6**, e25105, (2011).
29  Nickerl, J., Helbig, R., Schulz, H.-J., Werner, C. & Neinhuis, C. Diversity and potential correlations to the function of Collembola cuticle structures. *Zoomorphology* **132**, 183-195, (2013).
30  Noble-Nesbitt, J. Transpiration in Podura aquatica L.(Collembola, Isotomidae) and the wetting properties of its cuticle. *Journal of Experimental Biology* **40**, 681-700, (1963).
31  Marto, P. J. & Rohsenow, W. M. Nucleate Boiling Instability of Alkali Metals. *Journal of Heat Transfer* **88**, 183-&, (1966).
32  Webb, R. L. The evolution of enhanced surface geometries for nucleate boiling. *Heat Transfer Engineering* **2**, 46-69, (1981).
33  Kim, C. J. *Structured Surfaces for Enhanced Nucleate Boiling* MS thesis, Iowa State University, (1985).
34  Liu, T. & Kim, C.-J. Doubly re-entrant cavities to sustain boiling nucleation in FC-72. *2015 28th IEEE International Conference on Micro Electro Mechanical Systems (MEMS)*, 1122-1124, (2015).
35  Cassie, A. B. D. & Baxter, S. Wettability of porous surfaces. *Transactions of the Faraday Society* **40**, 0546-0550, (1944).
36  Wenzel, R. N. Resistance of solid surface to wetting by water. *Industrial and Engineering Chemistry* **28**, 7, (1936).
37  Arunachalam, S., Das, R., Nauruzbayeva, J., Domingues, E. M. & Mishra, H. Assessing omniphobicity by immersion. *Journal of Colloid and Interface Science* **534**, 156-162, (2019).
38  Domingues, E. M., Arunachalam, S. & Mishra, H. Doubly Reentrant Cavities Prevent Catastrophic Wetting Transitions on Intrinsically Wetting Surfaces. *ACS Applied Materials & Interfaces* **9**, 21532-21538, (2017).
39  Domingues, E. M., Arunachalam, S., Nauruzbayeva, J. & Mishra, H. Biomimetic coating-free surfaces for long-term entrapment of air under wetting liquids. *Nature Communications* **9**, 3606, (2018).
40  Mishra, H. *et al.* Time-Dependent Wetting Behavior of PDMS Surfaces with Bioinspired, Hierarchical Structures. *ACS Applied Materials & Interfaces* **8**, 8168-8174, (2016).

41   Kaufman, Y. *et al.* Simple-to-Apply Wetting Model to Predict Thermodynamically Stable and Metastable Contact Angles on Textured/Rough/Patterned Surfaces. *The Journal of Physical Chemistry C* **121**, 5642-5656, (2017).
42   Seo, D. *et al.* Rates of cavity filling by liquids. *Proceedings of the National Academy of Sciences*, (2018).
43   Hensel, R. *et al.* Wetting Resistance at Its Topographical Limit: The Benefit of Mushroom and Serif T Structures. *Langmuir* **29**, 1100-1112, (2013).
44   Hensel, R., Neinhuis, C. & Werner, C. The springtail cuticle as a blueprint for omniphobic surfaces. *Chemical Society Reviews* **45**, 323-341, (2016).
45   Spratlen, M. J. *et al.* The association between prenatal exposure to perfluoroalkyl substances and childhood neurodevelopment. *Environmental Pollution* **263**, 114444, (2020).
46   Seltenrich, N. From drinking water to individual body burden: bodeling toxicokinetics of four PFAS. *Environmental Health Perspectives* **131**, 014001, (2023).
47   Arunachalam, S. *et al.* Rendering SiO2/Si Surfaces Omniphobic by Carving Gas-Entrapping Microtextures Comprising Reentrant and Doubly Reentrant Cavities or Pillars. *JoVE (Journal of Visualized Experiments)*, e60403, (2020).
48   Das, R., Arunachalam, S., Ahmad, Z., Manalastas, E. & Mishra, H. Bio-inspired gas-entrapping membranes (GEMs) derived from common water-wet materials for green desalination. *Journal of Membrane Science*, 117185, (2019).
49   Mishra, H., Arunachalam, S. N. M., Domingues, E. M. & Das, R. Perfluorocarbon-free membranes for membrane distillation. *US Patent App. 17/056,809*, (2021).
50   Arunachalam, S., Ahmad, Z., Das, R. & Mishra, H. Counterintuitive Wetting Transitions in Doubly Reentrant Cavities as a Function of Surface Make-Up, Hydrostatic Pressure, and Cavity Aspect Ratio. *Advanced Materials Interfaces* **7**, 2001268, (2020).
51   Seo, J., García-Mayoral, R. & Mani, A. Pressure fluctuations and interfacial robustness in turbulent flows over superhydrophobic surfaces. *Journal of Fluid Mechanics* **783**, 448-473, (2015).
52   Luo, X.-w., Ji, B. & Tsujimoto, Y. A review of cavitation in hydraulic machinery. *Journal of Hydrodynamics* **28**, 335-358, (2016).
53   Wang, W., Lu, H. & Meng, G. Pressure fluctuation characteristics induced by cavitation in a centrifugal pump. *IOP Conference Series: Earth and Environmental Science* **163**, 012040, (2018).
54   Hashmi, A., Yu, G., Reilly-Collette, M., Heiman, G. & Xu, J. Oscillating bubbles: a versatile tool for lab on a chip applications. *Lab on a Chip* **12**, 4216-4227, (2012).
55   Dilip, D., Bobji, M. S. & Govardhan, R. N. Effect of absolute pressure on flow through a textured hydrophobic microchannel. *Microfluidics and Nanofluidics* **19**, 1409-1427, (2015).
56   Liu, Y. *et al.* Synergy of slippery surface and pulse flow: an anti-scaling solution for direct contact membrane distillation. *Journal of Membrane Science* **603**, 118035, (2020).
57   Kim, D., Hong, J. & Chung, S. K. Acoustic bubble array-induced jet flow for cleaning particulate contaminants on semiconductor wafers. *Korean Journal of Chemical Engineering*, 1-6, (2022).
58   Michelin, S., Guérin, E. & Lauga, E. Collective dissolution of microbubbles. *Physical Review Fluids* **3**, 043601, (2018).

59   Arunachalam, S., Mishra, H. Directional Wetting of Submerged Gas-entrapping Microtextures. *Under review*.
60   Iqtidar, A. *et al.* Drying dynamics of sessile-droplet arrays. *Physical Review Fluids* **8**, 013602, (2023).
61   Pleil, J. D., Wallace, M. A. G., Davis, M. D. & Matty, C. M. The physics of human breathing: Flow, timing, volume, and pressure parameters for normal, on-demand, and ventilator respiration. *Journal of breath research* **15**, 042002, (2021).
62   Piao, L. & Park, H. Two-dimensional analysis of air–water interface on superhydrophobic grooves under fluctuating water pressure. *Langmuir* **31**, 8022-8032, (2015).
63   Tabor, D. *Gases, Liquids and Solids and Other States of Matter*. (Cambridge University Press, 2000).
64   Bourgoun, A. & Ling, H. A General Model for the Longevity of Super-Hydrophobic Surfaces in Under-Saturated, Stationary Liquid. *Journal of Heat Transfer* **144**, 042101, (2022).
65   Patankar, N. A. Thermodynamics of trapping gases for underwater superhydrophobicity. *Langmuir* **32**, 7023-7028, (2016).
66   Lv, P., Xue, Y., Shi, Y., Lin, H. & Duan, H. Metastable states and wetting transition of submerged superhydrophobic structures. *Physical review letters* **112**, 196101, (2014).
67   Choi, W., Kang, M., Park, J. Y., Jeong, H. E. & Lee, S. J. Enhanced air stability of superhydrophobic surfaces with flexible overhangs of re-entrant structures. *Physics of Fluids* **33**, 022001, (2021).
68   Zhu, P. & Wang, L. Microfluidics-enabled soft manufacture of materials with tailorable wettability. *Chemical Reviews* **122**, 7010-7060, (2021).
69   Brennen, C. E. *Cavitation and bubble dynamics*. (Cambridge university press, 2014).
70   Yasui, K. Influence of ultrasonic frequency on multibubble sonoluminescence. *The Journal of the Acoustical Society of America* **112**, 1405-1413, (2002).
71   Crum, L. A. Measurements of the growth of air bubbles by rectified diffusion. *The Journal of the Acoustical Society of America* **68**, 203-211, (1980).
72   Wilke, K. L., Preston, D. J., Lu, Z. & Wang, E. N. Toward condensation-resistant omniphobic surfaces. *ACS nano* **12**, 11013-11021, (2018).
73   Lighton, J. R. Notes from underground: towards ultimate hypotheses of cyclic, discontinuous gas-exchange in tracheate arthropods. *American Zoologist* **38**, 483-491, (1998).
74   Hetz, S. K. & Bradley, T. J. Insects breathe discontinuously to avoid oxygen toxicity. *Nature* **433**, 516-519, (2005).
75   Crum, L. A. & Mao, Y. Acoustically enhanced bubble growth at low frequencies and its implications for human diver and marine mammal safety. *The Journal of the Acoustical Society of America* **99**, 2898-2907, (1996).
76   Shi, M., Das, R., Arunachalam, S. & Mishra, H. Suppression of Leidenfrost effect on superhydrophobic surfaces. *Physics of Fluids* **33**, 122104, (2021).
77   Zhang, J., Yao, Z. & Hao, P. Formation and evolution of air–water interfaces between hydrophilic structures in a microchannel. *Microfluidics and Nanofluidics* **21**, 135, (2017).
78   Pillai, S. *et al.* A Molecular to Macro-scale Assessment of Direct Contact Membrane Distillation for Separating Organics from Water. *Journal of Membrane Science*, (2020).
79   Bullee, P. A. Drag of wall-bounded flows. *PhD thesis, University of Twente*, (2020).

80   Liu, F.-W., Zhan, Y. & Cho, S. K. Effect of bubble interface position on propulsion and its control for oscillating-bubble powered microswimmer. *2018 IEEE Micro Electro Mechanical Systems (MEMS)*, 1173-1176, (2018).

**Acknowledgements**
The research work was supported by funding from King Abdullah University of Science and Technology (KAUST) under internal grant number BAS/1/1070-01-01. We thank Mr. Zain Ahmed for the fruitful technical discussions and Mr. Edelberto Manalastas for fabricating the chamber for pressure experiments. We also thank Prof. Wei Song Hwang (National University of Singapore) and Prof. Lanna Cheng (Scripps Institution of Oceanography) for kindly providing us with specimens of springtails whose images we have presented in Figure S1A-C. We also thank the operations team, Nanofabrication core lab, KAUST for microfabrication. Finally, we thank Mrs. Kate Zvorykina from Ella Maru studio for making the scientific illustrations.

**Competing interests**
The authors declare no competing interest.

**Data Availability.** All study data are included in the article and/or *SI*.

**Code availability**. The Mathematica code used in the current study are available from the corresponding author on reasonable request.

**Author Contributions:** S.A. and H.M. conceived the idea, S.A. performed the microfabrication, experiment, analysis and wrote the paper jointly with H.M. Theoretical modeling was performed by M.S.S.

# Supporting Information

**Coating-free Underwater Breathing via Biomimicry**

Sankara Arunachalam[1,2]*, Muhammad Subkhi Sadullah[1,2] & Himanshu Mishra[1,2,3]*

[1]Environmental Science and Engineering Program, Biological and Environmental Sciences and Engineering, Division, King Abdullah University of Science and Technology (KAUST), Thuwal 23955-6900, Kingdom of Saudi Arabia

[2]Water Desalination and Reuse Center, King Abdullah University of Science and Technology

[3]Center for Desert Agriculture, King Abdullah University of Science and Technology

*Corresponding authors: sankara.arunachalam@kaust.edu.sa & himanshu.mishra@kaust.edu.sa

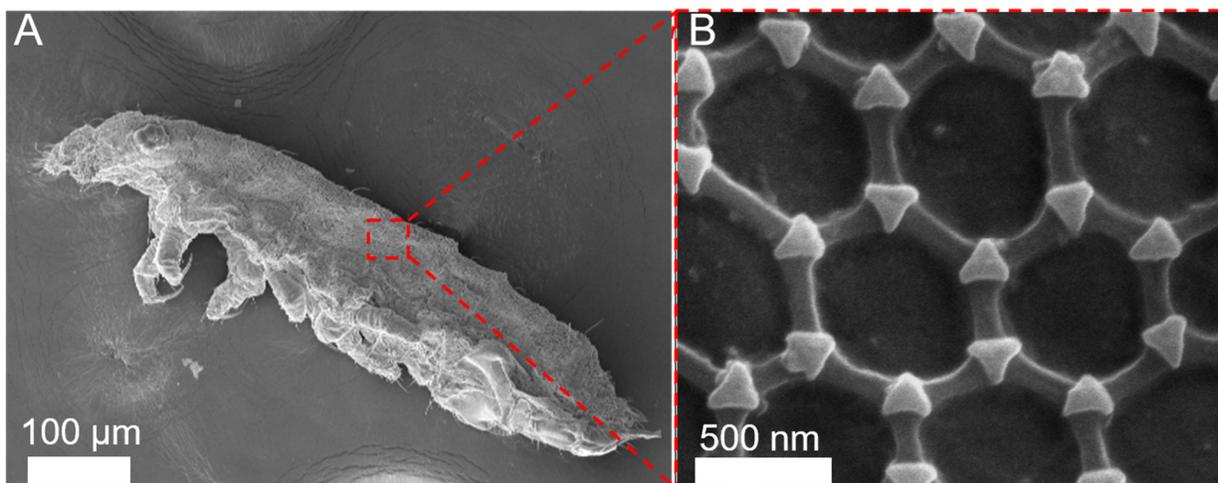

**Figure S1.** Representative scanning electron micrographs of the cuticles of springtails (Collembola) are covered by a layer of nanoscopic interconnected granules that form a basic hexagonal-like pattern.

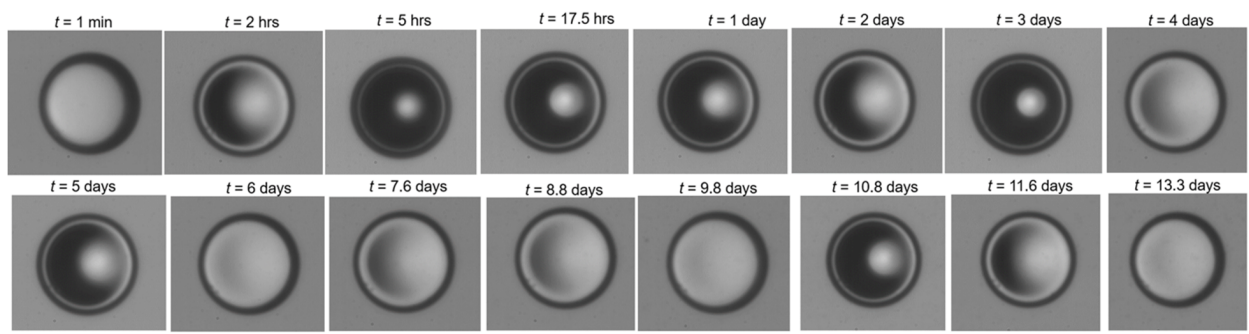

**Figure S2.** Optical micrographs (top-view) of silica surfaces with circular doubly reentrant cavity immersed in a 5 mm water column. The intrinsic contact angle on flat silica in air, $\theta_o \approx 63°$. No failure until 13 days. The diameter of the cavity is 200 µm, and the depth is 67 µm.

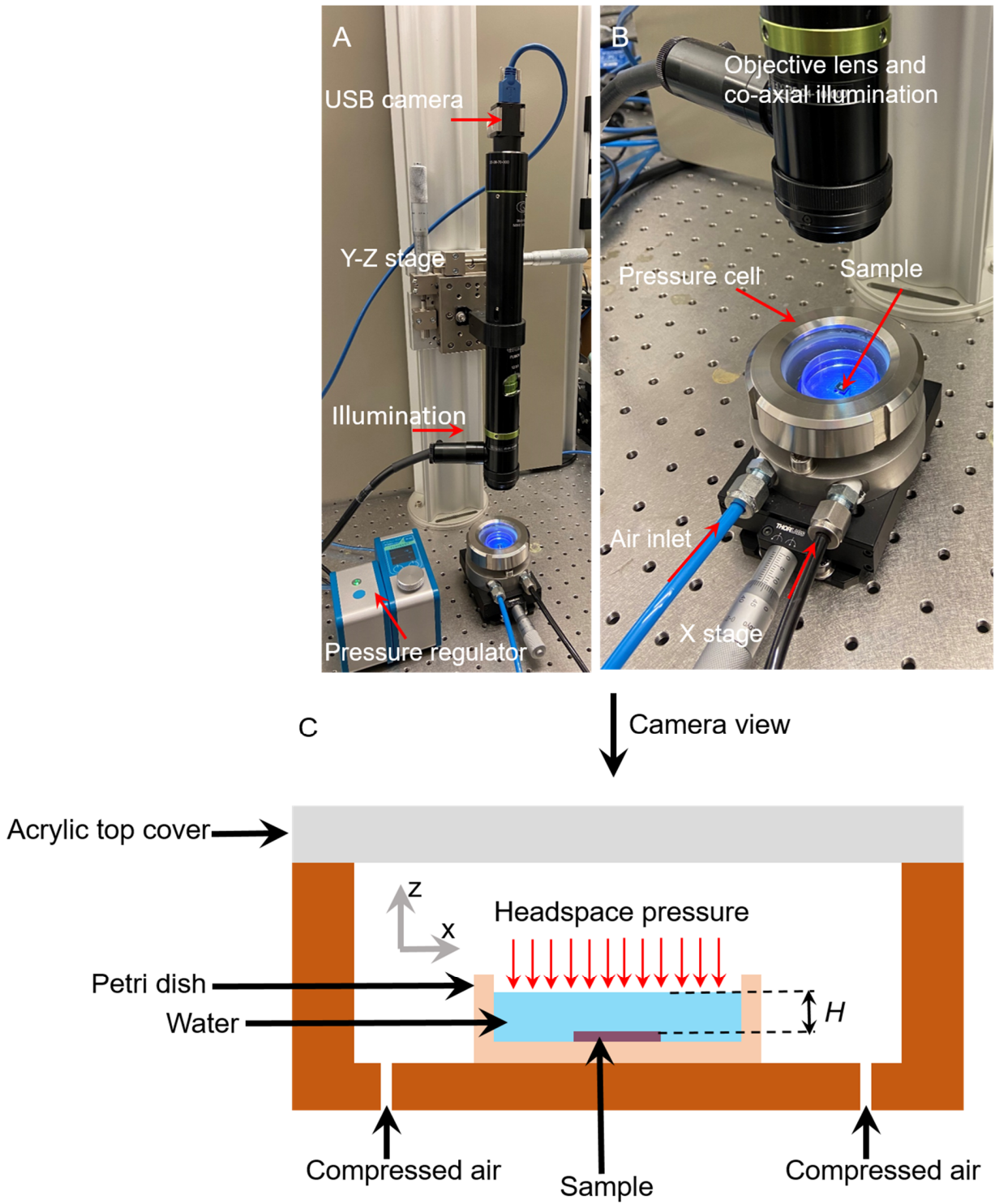

**Figure S3.** Experimental setup to determine the breakthrough pressures of cavities with water. Optical images were recorded using a vertically mounted microscope attached to a USB camera. **(B)** Samples were immersed in water inside the pressure cell made of stainless steel and polymethylmethacrylate top cover. Subsequently, we used compressed air to apply external pressure controlled by a fluigent pressure regulator. (C) Schematic of the pressure cell, where *H* is the water column height above the sample.

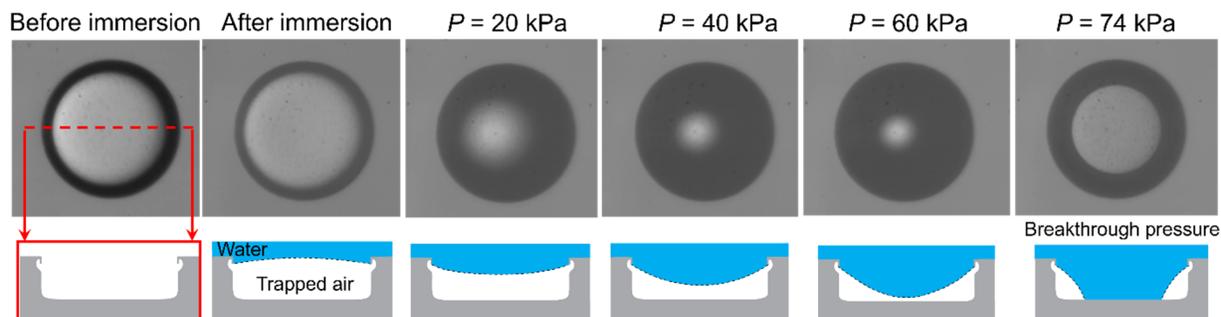

**Figure S4.** Breakthrough pressure was characterized under 3 mm-thick water columns by ramping up the pressure gradually at the rate of 1 kPa/s. The top panel shows the optical micrographs (top-view) of wetting transitions in a single doubly reentrant cavity under increasing hydrostatic pressure. The bottom panel shows the gas-liquid interface inside the cavity under increasing pressure (Scale bar: diameter of the cavity 200 µm). For the cavity depth $h = 55$ µm ($h = 67$ µm), the BtP was found to be 73.5 kPa (79 kPa).

**Section S1: Confocal Imaging**

Towards high-resolution visualization of the air-water interface under pressurization and depressurization, we built a custom-built pressure device to utilize confocal laser scanning microscopy (CLSM). After placing the sample's petri dish in the pressure cell, a syringe was used to gently pour water (containing 0.01 M Rhodamine B dye) until the sample was completely submerged. To minimize localized heating, we kept the laser intensity as low as possible (0.2 mW) and captured sequential line scan images in the Z-stack mode along the diameter of the cavity. The Line scan mode was used instead of the 2D scan to acquire images faster with minimal heating.

In a representative positive pressure cycling experiment, we raised the pressure to 20 kPa with a ramp rate of 1 kPa/s, held it for one minute for imaging, and then depressurized it followed by a three-minute interval time, and after a few cycles, the pressure was increased to 40 kPa(Figure S5 and S6). When DRC was immersed in water, i.e., with no applied pressure, the intruding air-water interface was slightly convex at the DR edge in the beginning (Figure S6A). Subsequently, with the application of the external pressure, from 0 to 20 kPa, the air-water interface started to become flat and sagging inside (concave) the cavity (Figure S6B). Eventually, as the pressure was relieved back to the ambient, the air-water meniscus began to move upward and reached a flat position, confirming a significant diffusional loss of air from the cavity (Figure S6C). With the higher number of cycles, the diffusional air loss from the cavity increased, which is evident from the meniscus shape (Figure S6D-F), and with the higher amplitude pressure cycle (40 kPa), the diffusional loss was more and the meniscus touched the cavity floor (failed, Figure S6G-L). Interestingly, throughout these wetting transitions, the water meniscus in contact with the DR edge at the inlet of the DRC remains pinned.

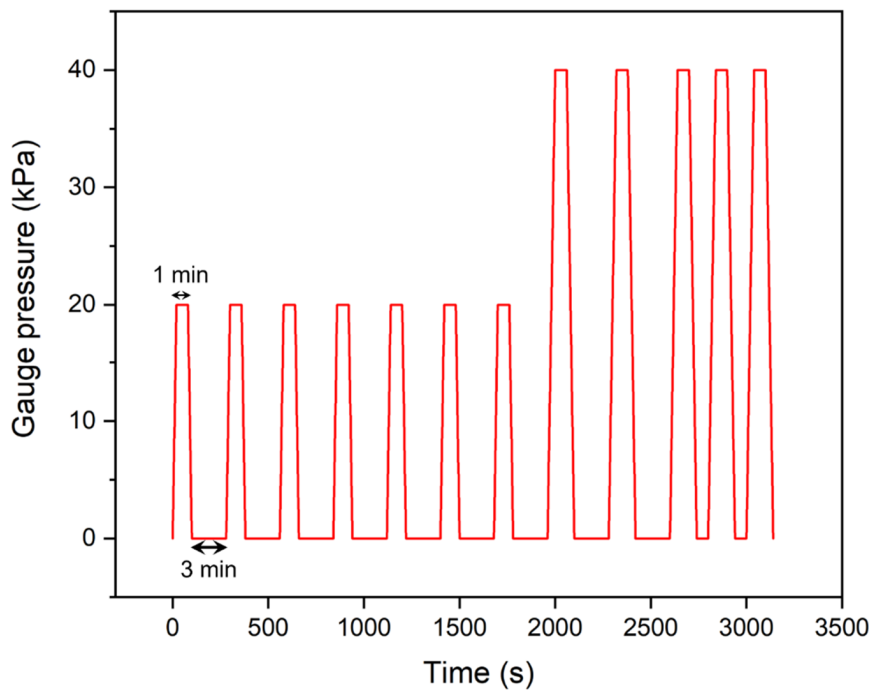

**Figure S5.** The cyclic pressure pattern used for the confocal imaging, presented in Fig. S6 A-L

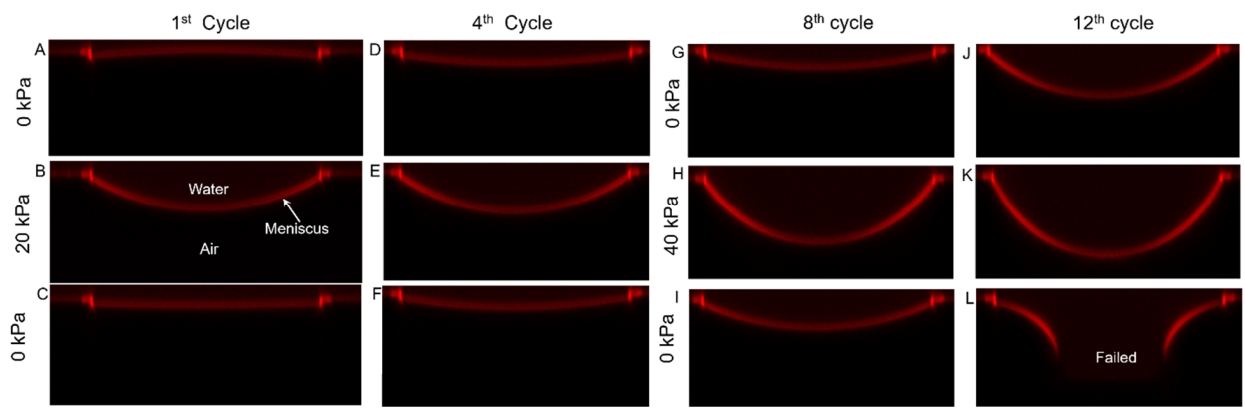

**Figure S6.** Confocal micrographs of wetting transitions of silica surfaces with a doubly reentrant cavity. 20 kPa amplitude cyclic pressure (A-F) followed by 40 kPa amplitude cyclic pressure (G-L) as shown in Fig. S5. We used line scan mode along with Z-stack to image across the diameter of the cavity (Scale bar: Cavity diameter 200 μm)

Here, we would like to point out that the heat from the CLSM-laser can accelerate wetting transitions with water, so only qualitative trends can be drawn[1,2]. In response, we utilized upright optical microscopy that does not suffer from these artifacts but lacks in resolution and visualization. Thus, CLSM and upright optical microscopy are quite complementary in investigating wetting transitions.

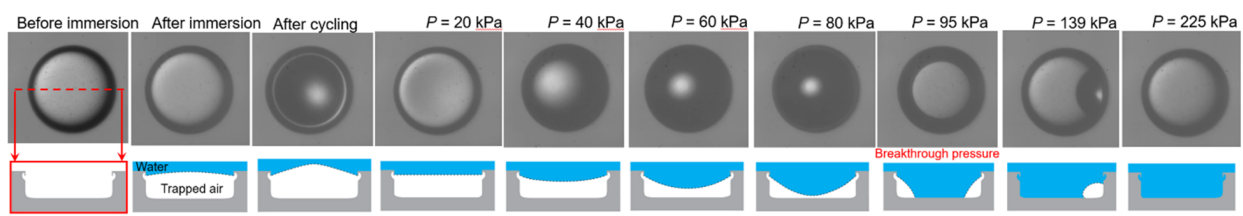

**Figure S7.** Enhancement in the breakthrough pressure after the cyclic pressure experiment. After 700 cycles of 40 kPa pressure amplitude with 10 min-long intercycle time, the breakthrough pressure increased by 28%. Water is oversaturated compared to the initial conditions. (Scale bar: diameter of the cavity, $D = 200$ μm, depth, $h = 55$ μm).

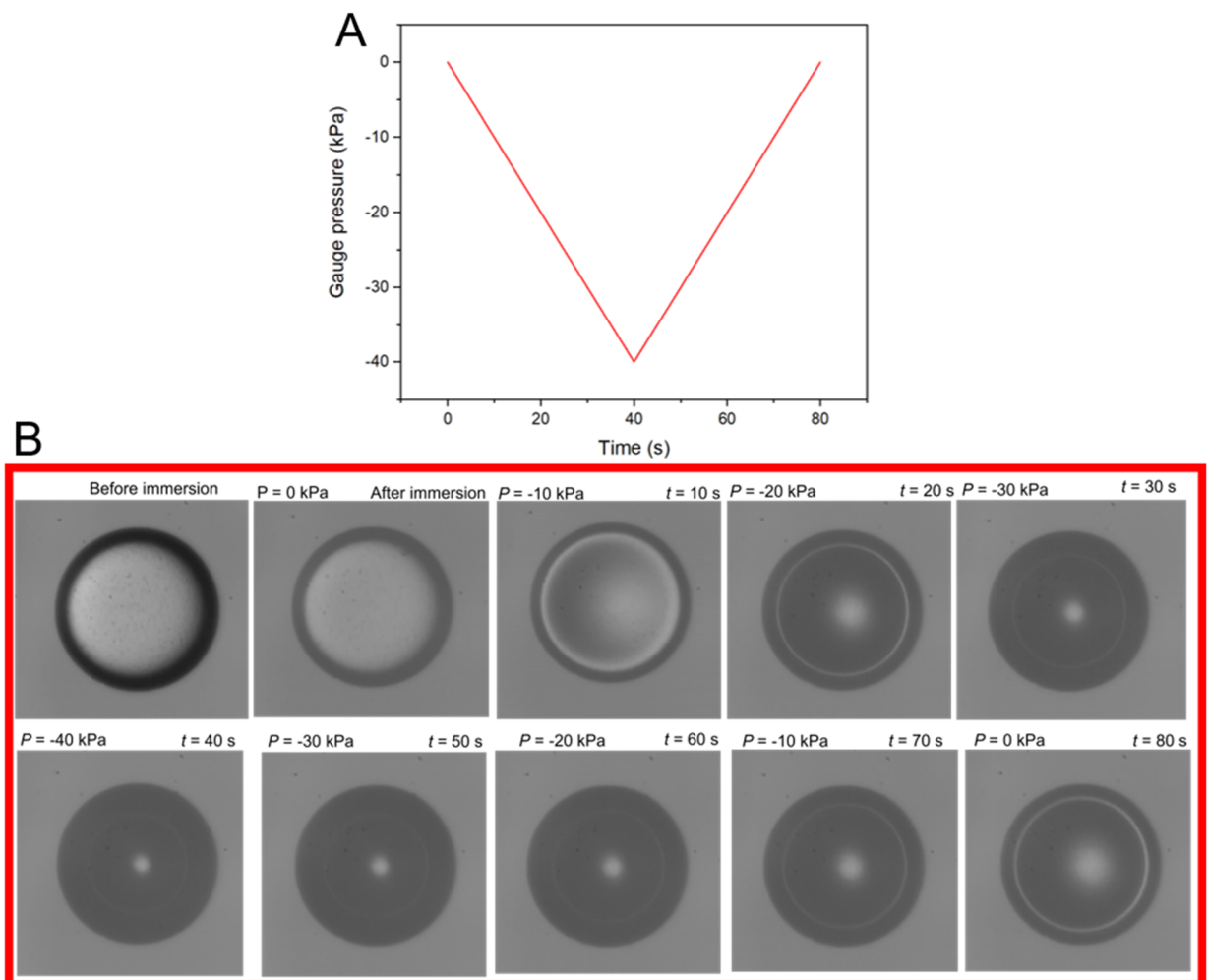

**Figure S8.** Top view optical micrographs of single -40 kPa pressure cycle, where headspace pressure decreases from 101 kPa to 61 kPa and recovers back to 101 kPa (1 atm) at 1 kPa/s. At the end of a single cycle, the air–water interface was found to be bulging upwards (convex).

**Section S2: Theoretical model**

In this section, we develop a theoretical model to predict the breakthrough pressure and failing cycle observed under our experimental conditions.

**Initial conditions**

The simulation works by tracking the gas volume inside the cavity under the effect of pressure change and diffusion. The initial gas volume at the immersion is given by $\pi D^2 (h - h_{DR})/4$, where $D$ and $h$ are the cavity diameter and depth (see Fig. 6A in the main text). The gas pressure before the immersion is $(P_{atm} - P_v)$, where $P_{atm}$ = 101.325 kPa is the atmospheric pressure and $P_v$ = 2.64 kPa is the vapor pressure of water. Immediately after the immersion, the meniscus of the contact angle $\theta_0$ is formed, and thus, the gas volume inside the cavity changes. Consequently, the gas pressure inside the cavity after the immersion $p_{G,0}$ also changes

$$p_{G,0} = (P_{atm} - P_v) \frac{\pi D^2 (h - h_{DR})/4}{V(\theta_0)}, \tag{S1}$$

where $V(\theta_0)$ is the gas volume after the immersion. In general, assuming that the meniscus is pinned at the edge of reentrant overhang with overhang length $h_{DR}$, the gas volume as a function of the interfacial angle can be expressed as

$$V(\theta) = \frac{\pi D^2}{4}(h - h_{DR}) + \frac{\pi D^3}{8} \frac{(2 - 3\sin\theta + \sin^3\theta)}{(3\cos^3\theta)}. \tag{S2}$$

Using this relation, we can probe the gas volume inside the cavity by means of $\theta$. This is practically easier than directly tracking the gas volume. Let us now define the initial external pressure as $P_i = (P_{atm} + \rho g H - P_v)$. Here, $\rho$, $g$, and $H$ are, respectively, the water density, gravitational acceleration, and thickness of the water column. Then, the initial condition can be described using the Laplace's law $\Delta P = 2\gamma/R$, where $\Delta P$ is the pressure difference across the interface given by $P_i - p_{G,0}$, $\gamma$ is the liquid surface tension and $R = -D\cos\theta/2$ is the curvature radius. Using the above definitions, we can write the initial condition as a function of the initial interfacial angle $\theta_0$ as

$$(P_{atm} + \rho g H - P_v) - \frac{(P_{atm} - P_v)\pi D^2 (h - h_{DR})/4}{\frac{\pi D^2}{4}(h - h_{DR}) + \frac{\pi D^3}{8}\frac{(2 - 3\sin\theta_0 + \sin^3\theta_0)}{(3\cos^3\theta_0)}} = -\frac{4\gamma\cos\theta_0}{D}. \tag{S3}$$

Here we have also assumed that there is no hydrostatic variation at the curvature since the cavity size is smaller than the capillary length. Eq. (S3) can be solved numerically using Mathematica software to obtain $\theta_0$. Then, the initial gas volume can be calculated from Eq. (S2) as $V_i = V(\theta_0)$.

**The evolution of the interfacial angle**

In the experiment, the pressure in the system is increased at a specific pressure ramp rate. We simulate this by iteratively increasing the external pressure by pressure increment $P$ at time step $dt$ for $N$ iterations. Using Eq. (S2) and $P_1 V_1 = P_2 V_2$ relation, we can calculate the gas volume and the interfacial angle after the pressure increase $\theta_p$ as

$$\frac{P_1 V_1}{P_2} = \frac{\pi D^2}{4}(h - h_{DR}) + \frac{\pi D^3}{8}\frac{(2 - 3\sin\theta_p + \sin^3\theta_p)}{(3\cos^3\theta_p)}. \tag{S4}$$

At the first iteration, $P_1 = P_i$, $V_1 = V_i$, and $P_2 = P_i + P$. Again, we can numerically solve Eq. (S4) to obtain $\theta_p$.

In addition, to describe the gas volume change due to the diffusional loss we used Fick's first law $J = -D_G \frac{(c_0 - c_\infty)}{l}$ along with Henry law, the ideal gas equation to obtain the following governing equation as discussed in the main manuscript

$$\frac{1}{A(\theta(t))}\frac{d[p_G V(\theta(t))]}{dt} = -RT\frac{D_G}{K_H}\frac{(p_G - s \cdot p_{G,0})}{l}, \tag{S5}$$

where $p_G$ is the current gas pressure, in this case $p_G = P_i + P$. The value of the Henry's constant and diffusion coefficient for air are respectively $K_H = 1.30 \times 10^5$ m³Pa mol⁻¹ and $D_G = 2\times 10^{-5}$ cm²/s. The interfacial area $A$ can also be expressed as a function of $\theta$ as follows

$$A(\theta) = \frac{\pi D^2}{(\cos(\theta/2) + \sin(\theta/2))^2}. \tag{S6}$$

We have also used $s \sim 1$ while $l$ is in the order of $10^{-4}$ m. By substituting Eq. (S2) and (S6) to Eq. (S5), we obtain the differential equation that describes the evolution of $\theta(t)$. The differential equation is then solved numerically for a small time step from $0 \geq t \geq \Delta t$ and using $\theta(0) = \theta_p$ as the initial value. The final $\theta$ value for this iteration is obtained by evaluating $\theta(\Delta t)$. The corresponding gas volume and pressure of this step is then become the initial values for the next iteration, i.e. $P_1 = p_G$, $V_1 = V(\theta(\Delta t))$, and $P_2 = p_G + P$.

**Simulation procedures**

We performed different simulation procedures depending on the type of experiment we want to replicate, e.g., continuous pressure ramps, cyclic pressures, and cyclic pressures with time interval. For the continuous pressure ramp $P_{\text{rate}}$, $P$ is increased by 0.1 kPa at every time step $\Delta t$ where it is determined by $\Delta t = P/P_{rate}$. The same also applies for the cyclic pressures, except once the iteration reaches $N = P_{max}/P$, $P$ and $P_{rate}$ become negative and the iteration continues until $p_G = P_I$ and complete one cycle. In addition, when the time interval $t_i$ is used, Eq. (S5) is solved from $0 \geq t \geq t_i$ after the completion of each cycle. These procedures were performed until the maximum contact angle $\theta_{max}$ was achieved which indicates that the interface has touched the bottom of the cavity. The value $\theta_{max}$ of is obtained from the following geometric relation

$$\frac{h - h_{DR}}{\sin \theta_{max} - 1} = \frac{D}{2 \cos \theta_{max}}. \tag{S7}$$

There are some differences between the model and the experimental observations (Fig. 6). This is expected since we only consider the diffusion in one dimension in our model. However, our simulation results show that our model can capture the experimental results, confirming that the governing physics is diffusion.

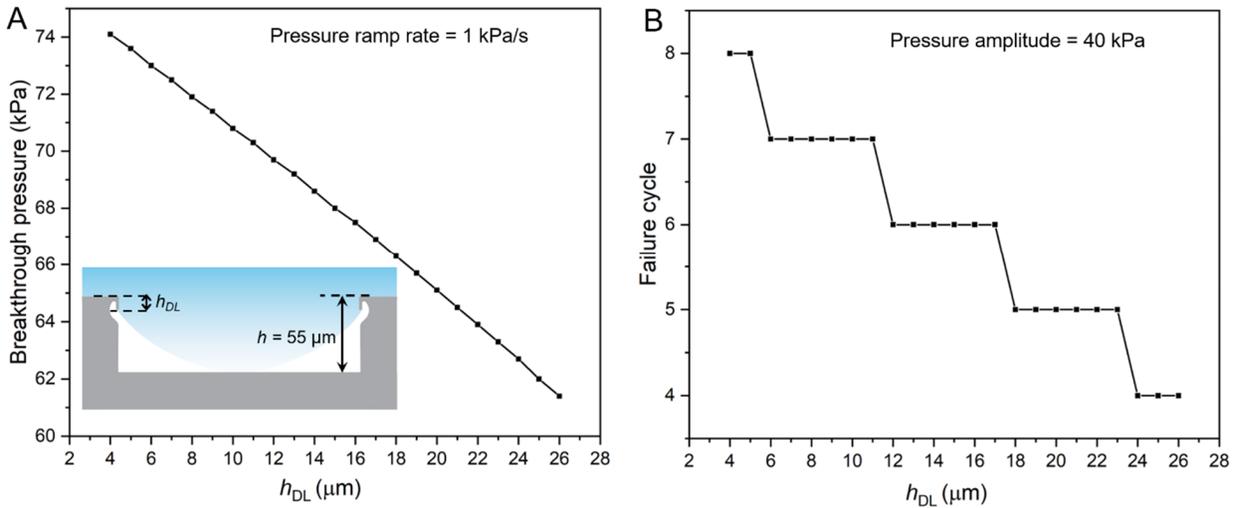

**Figure S9.** The influence of the length of the doubly reentrant overhang on the breakthrough pressure and the failure cycle number. (A-B) For a fixed cavity depth ($h$ =55 µm), increasing the $h_{DL}$ reduces the breakthrough pressure and the failure cycle number. As presented in Fig. 6 and panel (A), the longer the reentrant overhang ($h_{DL}$) the lower the $\theta_{max}$ and hence the lower the BtP. The stepwise decrease in the failure cycle number does not capture the quantitative variation in the Btp shown in Ain. Although for some $h_{DL}$ values have the same failing cycles, they actually failed at different BtP values with small gradual differences.

**Movie captions:**

**Movie S1**: The breakthrough pressure of doubly reentrant cavity. Top-view of microtextured silica surface with single doubly reentrant cavity immersed under 3 mm water column and headspace air pressure increased at 1 kPa/s until interface touches the cavity floor. Movie playback speed: 5x faster

**Movie S2**: The monotonically depleting air pocket inside the doubly reentrant cavity under cyclic pressure (no-interval). Top-view of microtextured silica surface with single doubly reentrant cavity immersed under 3 mm water column and 40 kPa amplitude cyclic pressure is applied continuously at 1kP/s. Movie playback speed: 48x faster

**Movie S3:** Indefinitely stable air pocket inside the doubly reentrant cavity under cyclic pressure with intervals time. Top-view of microtextured silica surface with single doubly reentrant cavity immersed under 3 mm water column. A 40 kPa amplitude cyclic pressure is applied at 1 kPa/s , and the interval between cycles varied 4, 5, 6 and 8 min. Indefinitely stable air pocket is observed for 6 and 8 min intervals. Movie playback speed: 1320x faster

**Movie S4**: The stability of the air pocket inside the doubly reentrant cavity subjected to cyclic pressure pattern similar to human breathing. Top-view of microtextured silica surface with single doubly reentrant cavity immersed under 6 and 2 mm water column. The breathe-in pressure was set to 8 kPa, the breathe-out pressure was set to 3.5 kPa, the breathing duration (in/out) was set to 1.5 sec, and the time interval between each cycle was 0 s. An indefinitely stable air pocket is observed for a 2 mm water column. Movie playback speed: 2200x faster

**Movie S5**: The air pocket growing outside the doubly reentrant cavity under negative pressure cycles (no-interval). Top-view of microtextured silica surface with single doubly reentrant cavity immersed under 3 mm water column and -40 kPa amplitude cyclic pressure is applied continuously at 1kP/s using vacuum suction. Movie playback speed: 2200x faster

**Movie S6**: The stable interface after bubble departure from the doubly reentrant cavity due to mechanical vibration. The air pocket grows outside the doubly reentrant cavity due to -40 kPa

amplitude cyclic pressure applied using vacuum suction. After a significant portion of the bubble departed due to mechanical vibration, the DRCs recovered their initial Cassie state. Movie playback speed: 1x

**Movie S7**: The air pocket growing outside the doubly reentrant cavity under positive-negative pressure cycles (no-interval). DRCs immersed in 3 mm-thick water columns was subjected to alternating positive–negative pressure cycles of amplitude ±40 kPa (at the ramp rate of ±1kPa/s and $t_i = 0$) via a combination of positive air pressure and vacuum suction. The trapped air bubble grew in its size over time. Movie playback speed: 4200x faster

References:

1    Lv, P. Y. *et al.* Symmetric and Asymmetric Meniscus Collapse in Wetting Transition on Submerged Structured Surfaces. *Langmuir* **31**, 1248-1254, (2015).
2    Domingues, E. M., Arunachalam, S., Nauruzbayeva, J. & Mishra, H. Biomimetic coating-free surfaces for long-term entrapment of air under wetting liquids. *Nature Communications* **9**, 3606, (2018).